\documentclass[preprint,aps]{revtex4}
\usepackage[dvips]{graphicx}

\begin{document}

\newcommand{\dfrac}[2]{\frac{\displaystyle #1}{\displaystyle #2}}
\newcommand{\nubar}{\chi}
\newcommand{\heavynu}{N}
\newcommand{\lightnu}{n}
\preprint{VPI--IPPAP--03--01}

\title{Quark-Lepton Unification and Lepton Flavor Non-Conservation
from a TeV-scale Seesaw Neutrino Mass Texture}
\author{Will~Loinaz\footnote{Electronic address: loinaz@alumni.princeton.edu}}
\affiliation{Department of Physics, Amherst College, Amherst MA 01002}
\author{Naotoshi~Okamura\footnote{Electronic address: nokamura@vt.edu},
Saifuddin~Rayyan\footnote{Electronic address: srayyan@vt.edu},
Tatsu~Takeuchi\footnote{Electronic address: takeuchi@vt.edu}}
\affiliation{Institute for Particle Physics and Astrophysics,
Physics Department, Virginia Tech, Blacksburg, VA 24061}
\author{L.~C.~R.~Wijewardhana\footnote{Electronic address: 
rohana@physics.uc.edu}}
\affiliation{Department of Physics, University of Cincinnati, 
Cincinnati OH 45221-0011}

\date{March 31, 2003}

\begin{abstract}
In a recent paper, we pointed out that mixing
of the light neutrinos with heavy gauge singlet states could
reconcile the $Z$-pole data from $e^+e^-$ colliders and the
$\nu_\mu$ ($\bar{\nu}_\mu$) scattering data from the NuTeV 
experiment at Fermilab.
We further noted that the mixing angle required to fit the data is much
larger than what would be expected from the conventional seesaw mechanism.
In this paper, we show how such mixings can be arranged
by a judicious choice of the neutrino mass texture.
We also argue that by invoking the unification of the Dirac
mass matrix for the up-type quarks and the neutrinos,
the mass of the heavy states can naturally be expected to lie
in the few TeV range.
The model is strongly constrained by the lepton flavor
changing process $\mu\rightarrow e\gamma$ 
which requires lepton universality
to be violated in the charged channel.
\end{abstract}


\maketitle

\section{Introduction}

In a previous paper \cite{LOTW1}, we argued that the
$Z$-pole data from $e^+e^-$ colliders \cite{LEP/SLD:2002}
and the $\nu_\mu$ ($\bar{\nu}_\mu$) scattering data from
the NuTeV experiment at Fermilab \cite{Zeller:2001hh}
could be brought into agreement if the Higgs boson is heavy, and if
the $Z\nu_\ell\nu_\ell$ couplings are suppressed by a factor
of $(1-\varepsilon_\ell)$ while the $W\ell\nu_\ell$ couplings are suppressed
by a corresponding factor of $(1-\varepsilon_\ell/2)$, 
where $\ell=e,\mu,\tau$.
Assuming lepton flavor universality of the suppression parameters
$\varepsilon_e = \varepsilon_\mu = \varepsilon_\tau \equiv \varepsilon$, 
the value of $\varepsilon$ required to fit the data is
\begin{equation}
\varepsilon = 0.0030 \pm  0.0010 \;.
\label{epslimit}
\end{equation}
Such a suppression could result if the neutrinos mix with 
heavy gauge singlet states.  However, the mixing angle required in 
the case of mixing with a single heavy state, 
\begin{equation}
\theta = 0.055 \pm 0.010\; , 
\end{equation}
is much larger than would be expected from a conventional seesaw model \cite{seesaw}.
Explicitly, the seesaw model relates the mass eigenvalues and 
the mixing angle by
\begin{equation}
\dfrac{ m_\mathrm{light} }{ m_\mathrm{heavy} }
\approx \theta^2\;.
\end{equation}
Reasonable values of the masses,
\begin{equation}
m_\mathrm{light} \sim 0.1\,\mathrm{eV}\;,\qquad
m_\mathrm{heavy} \sim 100\,\mathrm{GeV}\;,
\end{equation}
result in a mixing angle
\begin{equation}
\theta \sim 10^{-6}\;,
\end{equation}
which is too small by orders of magnitude.

This difficulty is absent in models which introduce more than one 
sterile neutrino per active flavor \cite{Gronau:ct}.
However, even in models with the same number of sterile neutrinos
as active neutrinos, Chang, Ng, and Ng~\cite{Chang:1994hz} have shown
that the seesaw constraint can be circumvented by allowing mixing among generations:
there are sufficient degrees of freedom in the neutrino mixing matrix that
the masses of the light and heavy neutrino states can be tuned independently of
their mixing angles.

In the following, we will present explicit examples of mass matrices
(textures) that demonstrate this feature.  
We show that the light neutrinos can be
made massless while maintaining large mixing angles with
the heavy states, the masses of which can also be adjusted at will.
We also derive the relation between the suppression parameter 
$\varepsilon$, mixing angles,
and the mass parameters in the textures.

A natural question to ask is: what are the masses of the heavy states in these models?
Since the mixing angles are dimensionless, they fix only certain ratios 
between the mass parameters in the texture.
The masses of the heavy states are therefore, in principle, unconstrained.
Quark--lepton unification, as appears, for example, in 
the Pati--Salam model \cite{Pati:uk}, can fix the order of the
Dirac masses in the texture to be $\sim 100$~GeV.
In combination with the experimental limits on the suppression parameter,
Eq.~(\ref{epslimit}), this allows the
heavy state masses to be of order a few TeV.
Such states may be directly observable at the
Large Hadron Collider (LHC), currently under construction at CERN.  

The effect of the heavy states on the flavor--changing process
$\mu\rightarrow e\gamma$ 
\cite{Chang:1994hz,mutoegammaold,mutoegamma,mutoegammarecent}
places a strong constraint on the model.  
The upper bound on the branching fraction 
\begin{equation}
B(\mu\rightarrow e\gamma) = 
\dfrac{\Gamma(\mu\rightarrow e\gamma)}
      {\Gamma(\mu\rightarrow e\nu\bar{\nu})}\;,
\end{equation}
from the MEGA collaboration \cite{mega} requires that the product 
$\varepsilon_e\varepsilon_\mu$ be strongly suppressed.
Since either $\varepsilon_e$ or $\varepsilon_\mu$ must remain sizable
to account for the discrepancy between the $Z$-pole and NuTeV data,
the violation of lepton flavor universality among the 
$\varepsilon_\ell$'s is required.

Finally, we compare our results to those of a recent analysis of 
our proposal by Glashow \cite{Glashow:2003wt}.


\section{One Generation Case}

First, let us trace the problem with the seesaw mechanism to its origin.
We assume the same number of $SU(2)_L$ active neutrinos ($\nu$) and 
sterile neutrinos ($\nubar$).
We further adopt the `seesaw' Ansatz in which the Majorana masses of 
the active neutrinos are all set to zero.
For a single generation, the seesaw mass matrix can only be
\begin{equation}
\left[ \; \nu \quad \nubar \; \right]
\left[ \begin{array}{cc} 0 & m \\
                          m & M
        \end{array}
\right]
\left[ \begin{array}{c} \nu \\ \nubar
        \end{array}
\right]\;.
\end{equation}
Note that there are only two free parameters: $m$ and $M$.
This mass matrix can be diagonalized with the orthogonal transformation 
$O$:
\begin{equation}
O^\mathrm{T}
\left[ \begin{array}{cc} 0 & m \\ m & M \end{array}
\right]
O =
\left[ \begin{array}{cc} - m\,t & 0 \\ 0 & m/t \end{array}
\right]\;,
\end{equation}
in which
\begin{equation}
O =
\left[ \begin{array}{cc} c & s \\ -s & c \end{array}
\right]\;,
\end{equation}
and
\begin{equation}
s = \sin\theta\;,\quad
c = \cos\theta\;,\quad
t = \tan\theta\;,\quad
\tan 2\theta = \frac{2m}{M}\;.
\label{onegen}
\end{equation}
Alternately, to make both mass eigenvalues positive we can diagonalize
using the unitary matrix $U$:
\begin{equation}
U =
\left[ \begin{array}{cc} ic & s \\ -is & c \end{array}
\right]
\end{equation}
which yields
\begin{equation}
U^\mathrm{T}
\left[ \begin{array}{cc} 0 & m \\ m & M \end{array}
\right]
U =
\left[ \begin{array}{cc} m\,t & 0 \\ 0 & m/t \end{array}
\right]\;.
\end{equation}
The product of the two mass eigenvalues is always equal to $m^2$ 
regardless the value of $M$.  Thus the nomenclature `seesaw model':
enhancing one of the mass eigenvalues suppresses the other.
In particular, when $m\ll M$ the two mass eigenvalues are
\begin{eqnarray}
m_\mathrm{light}
& \equiv & m\,t
\;=\; \frac{M}{2}\left( \sqrt{ 1 + \frac{4m^2}{M^2} } - 1 \right)
\;\approx\; \frac{m^2}{M}\;,  \cr
m_\mathrm{heavy}
& \equiv & \frac{m}{t}
\;=\; \frac{M}{2}\left( \sqrt{ 1 + \frac{4m^2}{M^2} } + 1 \right)
\;\approx\; M\;.
\end{eqnarray}
The mass eigenstates and the original generation eigenstates are related through
\begin{equation}
\left[ \begin{array}{c} \nu \\ \nubar \end{array}
\right] = U
\left[ \begin{array}{c} \lightnu \\ \heavynu
        \end{array}
\right] =
\left[ \begin{array}{rr} ic\,\lightnu + s\,\heavynu \\
                         -is\,\lightnu + c\,\heavynu
        \end{array}
\right]\;,
\end{equation}
where the light and heavy states are denoted $\lightnu$ and  
$\heavynu$, respectively.  
The suppression factor $(1-\varepsilon)$ is just the squared magnitude of
the coefficient of the light state $\lightnu$ in the active neutrino $\nu$,
\begin{equation}
1 - \varepsilon = \cos^2\theta
\approx 1 - \theta^2
\approx 1 - \frac{m^2}{M^2}\;.
\end{equation}
Observe that
\begin{equation}
\theta^2 \approx \frac{m^2}{M^2}
\approx \frac{ m_\mathrm{light} }{ m_\mathrm{heavy} }\;,
\end{equation}
which is the problem anticipated in the introduction.
This stems from the fact that
the original mass matrix had just two free parameters,  
so the two mass eigenvalues
$m_\mathrm{light} \sim m^2/M$, $m_\mathrm{heavy} \sim M$,
and the mixing angle $\theta\sim m/M$ are necessarily
related.

\section{Two Generation Case}

With the introduction of additional generations, the mass matrix has
more free parameters and the mixing angle and
the mass eigenvalues are in general independent.
Consider first the two generation case.
The most general seesaw mass matrix is
\begin{equation}
\left[ \begin{array}{cc} 
       \mathbf{0}_{2\times 2} & \mathbf{m}_{2\times 2} \\
       \mathbf{m}^\mathrm{T}_{2\times 2} & \mathbf{M}_{2\times 2}
       \end{array}
\right]\;,
\end{equation}
where $\mathbf{m}_{2\times 2}$ is the Dirac mass matrix which couples the active and
sterile neutrinos, and 
$\mathbf{M}_{2\times 2}$ is the symmetric Majorana mass matrix
for the steriles.
As a particular instance, take the following generation universal form:
\begin{equation}
\left[\;\nu_1\;\nu_2\;\nubar_1\;\nubar_2\;\right]
\left[ \begin{array}{rrrr}  0 &  0 & m & -m \\
                            0 &  0 & m & -m \\
                            m &  m & M &  0 \\
                           -m & -m & 0 & -M
       \end{array}
\right]
\left[ \begin{array}{c} \nu_1 \\ \nu_2 \\ \nubar_1 \\ \nubar_2
       \end{array}
\right]\;.
\label{twogentex}
\end{equation}
This mass matrix is manifestly rank two, so two of the mass eigenvalues will
be zero regardless the values of $m$ and $M$.
This matrix can be diagonalized as
\begin{equation}
O^\mathrm{T}
\left[ \begin{array}{rrrr}  0 &  0 & m & -m \\
                            0 &  0 & m & -m \\
                            m &  m & M &  0 \\
                           -m & -m & 0 & -M
       \end{array}
\right]
O =
\left[ \begin{array}{cccc} 0 & 0 & 0 & 0 \\
                           0 & 0 & 0 & 0 \\
                           0 & 0 & m/sc & 0 \\
                           0 & 0 & 0 & -m/sc
       \end{array}
\right]\;,
\end{equation}
in which 
\begin{equation}
O =
\left[ \begin{array}{rrrr} c^2 & -s^2 &  sc  &  sc  \\
                          -s^2 &  c^2 &  sc  &  sc  \\
                          -sc  & -sc  &  c^2 & -s^2 \\
                          -sc  & -sc  & -s^2 &  c^2
       \end{array}
\right]\;,
\label{twogenO}
\end{equation}
and
\begin{equation}
s = \sin\theta\;,\quad
c = \cos\theta\;,\quad
t = \tan\theta\;,\quad
\tan 2\theta = \frac{2m}{M}\;.
\label{twogen}
\end{equation}
The non-zero mass eigenvalues are
\begin{equation}
m_\mathrm{heavy} = \pm\frac{m}{sc} = \pm\sqrt{M^2+4m^2}\;.
\end{equation}
Note that the orthogonal matrix which diagonalizes 
our mass matrix is not unique since we are free to rotate
the zero mass eigenstates into each other.  
However, the above form is the one that reduces to the unit matrix in the limit 
$m/M\rightarrow 0$ and facilitates the extraction of the suppression factor.

The mass eigenstates and the generation eigenstates are related through
\begin{equation}
\left[ \begin{array}{c} \nu_1 \\ \nu_2 \\ \nubar_1 \\ \nubar_2
       \end{array}
\right] = O
\left[ \begin{array}{c}
          \lightnu_1 \\ \lightnu_2 \\
          \heavynu_1 \\ \heavynu_2
       \end{array}
\right] =
\left[ \begin{array}{r}
    c^2\,\lightnu_1 - s^2\,\lightnu_2
  + sc\,\heavynu_1 +sc\,\heavynu_2 \\
  - s^2\,\lightnu_1 + c^2\,\lightnu_2
  + sc\,\heavynu_1 +sc\,\heavynu_2 \\
  - sc\,\lightnu_1 - sc\,\lightnu_2
  + c^2\,\heavynu_1 - s^2\,\heavynu_2 \\
  -sc\,\lightnu_1 - sc\,\lightnu_2
  - s^2\,\heavynu_1 + c^2\,\heavynu_2
       \end{array}
\right]\;.
\end{equation}
The suppression factors for the $Z\nu_1\nu_1$ and $Z\nu_2\nu_2$ vertices 
can be read off from this expression:
\begin{equation}
1 - \varepsilon = \cos^4\theta
\approx 1 - 2\theta^2
\approx 1 - \frac{2m^2}{M^2}\;.
\label{twogensup}
\end{equation}
Thus, in contrast with the one generation case, the mass of the 
heavy states, $\sqrt{M^2+4m^2}$, and the suppression factor, 
$\varepsilon\approx 2m^2/M^2$,
can be adjusted independently since the two free parameters in this model, 
$m$ and $M$, are not constrained in any way by the light neutrino masses.

Of course, the light mass eigenstates are exactly massless
in this model so it is unrealistic as it stands.  
However, this can be remedied by 
adding a small perturbation to the mass matrix.  
For instance, consider the following perturbation:
\begin{equation}
\left[ \begin{array}{cccc}  0 &  0 & (1+\delta) m & (-1+\delta) m \\
                            0 &  0 & (1+\delta) m & (-1+\delta) m \\
                           (1+\delta) m &  (1+\delta) m & (1+\delta) M &  0 \\
                          (-1+\delta) m & (-1+\delta) m & 0 & (-1+\delta) M
       \end{array}
\right]\;,
\label{twogenpert}
\end{equation}
where $\delta\ll 1$.  This increases the rank of the matrix by one and renders
one of the massless states massive.  
Partial diagonalization can be obtained as
\begin{equation}
O^{\prime\mathrm{T}}O^\mathrm{T}
\left[ \begin{array}{cccc}  0 &  0 & (1+\delta) m & (-1+\delta) m \\
                            0 &  0 & (1+\delta) m & (-1+\delta) m \\
                           (1+\delta) m &  (1+\delta) m & (1+\delta) M &  0 \\
                          (-1+\delta) m & (-1+\delta) m & 0 & (-1+\delta) M
       \end{array}
\right]
O\,O' =
\left[ \begin{array}{cccc} 0 & 0 & 0 & 0 \\
                           0 &   &   &   \\
                           0 &   & \mathcal{M} & \\
                           0 &   &   &   
       \end{array}
\right]\;,
\end{equation}
where $O$ is the matrix given in Eq.~(\ref{twogenO}), and $O'$ is the matrix which
mixes the first two columns maximally,
\begin{equation}
O' =
\left[ \begin{array}{cccc}\phantom{-}\frac{1}{\sqrt{2}} &  \frac{1}{\sqrt{2}} &  0  &  0  \\
                          -\frac{1}{\sqrt{2}} &  \frac{1}{\sqrt{2}} &  0  &  0  \\
                          0  &  0  &  1  &  0 \\
                          0  &  0  &  0  &  1
       \end{array}
\right]\;.
\label{twogenOprime}
\end{equation}
The matrix $\mathcal{M}$ has the form
\begin{equation}
\mathcal{M} = M
\left[ \begin{array}{ccc} -4\delta\mu & -4\sqrt{2}\delta\mu r & -4\sqrt{2}\delta\mu r \\
                          -4\sqrt{2}\delta\mu r & \phantom{+}\sqrt{1+4r^2}+\delta(1+2\mu) & 2\delta\mu  \\
                          -4\sqrt{2}\delta\mu r & 2\delta\mu & -\sqrt{1+4r^2}+\delta(1+2\mu)
       \end{array}
\right]\;,
\end{equation}
where
\begin{equation}
r \equiv \frac{m}{M}\;,\qquad
\mu \equiv \frac{r^2}{1+4r^2} \;.
\label{rdef}
\end{equation}
Complete diagonalization of $\mathcal{M}$ requires further mixings by angles of order $\delta$.
In the limit $\delta\ll 1$, they can be neglected.

As another example, consider the perturbation
\begin{equation}
\left[ \begin{array}{cccc}  0 &  0 & m & -m \\
                            0 &  \delta M & (1-\delta) m & (-1-\delta) m \\
                            m &  (1-\delta) m & M &  0 \\
                           -m & (-1-\delta) m & 0 & -M
       \end{array}
\right]\;,
\label{twogenpert2}
\end{equation}
where we have allowed for a small ($\delta\ll 1$) but non-zero Majorana mass for one of the
active neutrinos.  This matrix can be partially diagonalized as
\begin{eqnarray}
\lefteqn{O^{\prime\prime\mathrm{T}}O^\mathrm{T}
\left[ \begin{array}{cccc}  0 &  0 & m & -m \\
                            0 &  \delta M & (1-\delta) m & -(1+\delta) m \\
                            m &  (1-\delta) m & M &  0 \\
                           -m & -(1+\delta) m & 0 & -M
       \end{array}
\right]
O\,O''} \cr 
& = & M
\left[ 
\begin{array}{cccc} 
-4\delta r^4 & 0 & 2\delta r^3 & 2\delta r^3 \\
0 & \delta\left( 1 + 2r^2 - 4r^4 \right) & 2\delta r^3 & 2\delta r^3 \\
2\delta r^3 & 2\delta r^3 & \sqrt{1+4r^2}-\delta\mu & -\delta\mu \\
2\delta r^3 & 2\delta r^3 & -\delta\mu & -\sqrt{1+4r^2}-\delta\mu \\
\end{array}
\right] + \mathcal{O}(\delta r^5) \;,
\end{eqnarray}
where $O$ is the matrix given in Eq.~(\ref{twogenO}), $O''$ is the matrix which
mixes the first two columns with angle $\theta'$,
\begin{equation}
O'' =
\left[ \begin{array}{cccc}\phantom{-}\cos\theta' & \sin\theta' &  0  &  0  \\
                          -\sin\theta' & \cos\theta' &  0  &  0  \\
                          0  &  0  &  1  &  0 \\
                          0  &  0  &  0  &  1
       \end{array}
\right]\;,\qquad
\tan 2\theta'\equiv 2c^2 s^2(c^2-s^2) \approx \frac{2m^2}{M^2}\;,
\label{twogenO2prime}
\end{equation}
and the parameters $r$ and $\mu$ are as defined in Eq.~(\ref{rdef}).
Again, complete diagonalization requires further mixings by angles of order $\delta$
which can be neglected.

As these examples illustrate, minute perturbations to the mass matrix, 
Eq.~(\ref{twogentex}), can generate a variety of masses and mixings 
for the light neutrinos.
Note that the extra mixings by $O'$ or $O''$ in these models have no effect on 
the suppression parameter, Eq.~(\ref{twogensup}), which remains common for 
$\nu_1$ and $\nu_2$.
This is to be expected as long as the perturbations are small and 
the mass matrix retains an approximate generation universal form.
However, that is not to say that the suppression 
is \textit{flavor} universal since the flavor eigenstates can be some mixture 
of $\nu_1$ and $\nu_2$ due to the mixing among the charged leptons. In general,
\begin{eqnarray}
\left[ \begin{array}{c} \nu_e \\ \nu_\mu 
       \end{array}
\right] 
& = & 
\mathcal{U}
\left[ \begin{array}{c} \nu_1 \\ \nu_2
       \end{array}
\right] 
\;=\;
\left[ \begin{array}{rr} \cos\phi & -\sin\phi \\
                         \sin\phi &  \cos\phi
       \end{array}
\right]
\left[ \begin{array}{c} \nu_1 \\ \nu_2
       \end{array}
\right] \cr
& = &
\left[ \begin{array}{c} (c^2\cos\phi + s^2\sin\phi) \lightnu_1
                       -(c^2\sin\phi + s^2\cos\phi) \lightnu_2 + \cdots \\
                        (c^2\sin\phi - s^2\cos\phi) \lightnu_1
                       +(c^2\cos\phi - s^2\sin\phi) \lightnu_2 + \cdots 
       \end{array}
\right] \;.
\label{twogenmix}
\end{eqnarray}
This mixing breaks the universality of the suppression factors.
For the $Z\nu_e\nu_e$ and $W e\nu_e$ vertices, the suppression factor is
\begin{eqnarray}
1 - \varepsilon_{e}
& = & (c^2\cos\phi + s^2\sin\phi)^2 + (s^2\cos\phi + c^2\sin\phi)^2 \cr
& = & c^4 + 2 s^2 c^2 \sin 2\phi + s^4 \cr
& \approx & 1 - 2\theta^2 ( 1 - \sin 2\phi ) \cr
& \approx & 1 - \varepsilon ( 1 - \sin 2\phi )\;, 
\end{eqnarray}
while that for the $Z\nu_\mu\nu_\mu$ and $W\mu\nu_\mu$ vertices is
\begin{eqnarray}
1 - \varepsilon_{\mu}
& = & (c^2\sin\phi - s^2\cos\phi)^2 + (c^2\cos\phi - s^2\sin\phi)^2 \cr
& = & c^4 - 2 s^2 c^2 \sin 2\phi + s^4 \cr
& \approx & 1 - 2\theta^2 ( 1 + \sin 2\phi ) \cr
& \approx & 1 - \varepsilon ( 1 + \sin 2\phi )\;.
\end{eqnarray}
Note that the sum $\varepsilon_e + \varepsilon_\mu$ is independent of the 
mixing angle $\phi$.
We will see later that this breaking of flavor universality is
important in satisfying the constraint from $\mu\rightarrow e\gamma$.

\section{Three Generation Case}

Extension of our model to the three generation case is straightforward.
The most general seesaw mass matrix for the three generation case is
\begin{equation}
\left[ \begin{array}{cc}
       \mathbf{0}_{3\times 3} & \mathbf{m}_{3\times 3} \\
       \mathbf{m}^\mathrm{T}_{3\times 3} & \mathbf{M}_{3\times 3}
       \end{array}
\right]\;,
\end{equation}
where $\mathbf{m}_{3\times 3}$ is the Dirac mass matrix, and
$\mathbf{M}_{3\times 3}$ is the symmetric Majorana mass matrix.
Consider the following Ansatz:
\begin{equation}
\left[\;\nu_1\;\nu_2\;\nu_3\;\nubar_1\;\nubar_2\;\nubar_3\;\right]
\left[ \begin{array}{cccccc}
        0 &  0 &  0 & \alpha m & \beta m & \gamma m \\
        0 &  0 &  0 & \alpha m & \beta m & \gamma m \\
        0 &  0 &  0 & \alpha m & \beta m & \gamma m \\
        \alpha m & \alpha m & \alpha m & \alpha M & 0 & 0 \\
        \beta  m & \beta  m & \beta  m & 0 & \beta  M & 0 \\
        \gamma m & \gamma m & \gamma m & 0 & 0 & \gamma M \\
        \end{array}
\right]
\left[ \begin{array}{c} \nu_1 \\ \nu_2 \\ \nu_3 \\
                         \nubar_1 \\ \nubar_2 \\ \nubar_3
        \end{array}
\right]\;,
\label{threegentex}
\end{equation}
where $\alpha + \beta + \gamma = 0$.
This matrix is manifestly at most rank three so that
at least three of the mass eigenvalues are zero.
This matrix can be block diagonalized as
\begin{equation}
O^\mathrm{T}
\left[ \begin{array}{cccccc}
        0 &  0 &  0 & \alpha m & \beta m & \gamma m \\
        0 &  0 &  0 & \alpha m & \beta m & \gamma m \\
        0 &  0 &  0 & \alpha m & \beta m & \gamma m \\
        \alpha m & \alpha m & \alpha m & \alpha M & 0 & 0 \\
        \beta  m & \beta  m & \beta  m & 0 & \beta  M & 0 \\
        \gamma m & \gamma m & \gamma m & 0 & 0 & \gamma M
        \end{array}
\right]
O =
\left[ \begin{array}{cc} 0 & 0 \\
                          0 & \mathcal{M}
        \end{array}
\right]\;,
\end{equation}
where
\begin{equation}
O =
\left[ \begin{array}{cccccc}
      1-\frac{2}{3}s^2 &  -\frac{2}{3}s^2 &  -\frac{2}{3}s^2 &
        \frac{2}{3}sc  &   \frac{2}{3}sc  &   \frac{2}{3}sc \\
       -\frac{2}{3}s^2 & 1-\frac{2}{3}s^2 &  -\frac{2}{3}s^2 &
        \frac{2}{3}sc  &   \frac{2}{3}sc  &   \frac{2}{3}sc \\
       -\frac{2}{3}s^2 &  -\frac{2}{3}s^2 & 1-\frac{2}{3}s^2 &
        \frac{2}{3}sc  &   \frac{2}{3}sc  &   \frac{2}{3}sc \\
       -\frac{2}{3}sc  &  -\frac{2}{3}sc  &  -\frac{2}{3}sc  &
      1-\frac{2}{3}s^2 &  -\frac{2}{3}s^2 &  -\frac{2}{3}s^2 \\
       -\frac{2}{3}sc  &  -\frac{2}{3}sc  &  -\frac{2}{3}sc  &
       -\frac{2}{3}s^2 & 1-\frac{2}{3}s^2 &  -\frac{2}{3}s^2 \\
       -\frac{2}{3}sc  &  -\frac{2}{3}sc  &  -\frac{2}{3}sc  &
       -\frac{2}{3}s^2 &  -\frac{2}{3}s^2 & 1-\frac{2}{3}s^2 \\
        \end{array}
\right]\;,
\end{equation}
\begin{equation}
\mathcal{M}
= \left[ \begin{array}{ccc}
          \alpha(M+2mt) & -\gamma m t   & -\beta  m t \\
          -\gamma m t   & \beta(M+2m t) & -\alpha m t \\
          -\beta  m t   & -\alpha m t   & \gamma(M+2m t)
          \end{array}
   \right]\;,
\end{equation}
and
\begin{equation}
s = \sin\theta\;,\quad
c = \cos\theta\;,\quad
t = \tan\theta\;,\quad
\tan 2\theta = \frac{3m}{M}\;.
\end{equation}
Note the difference in the definition of $\tan 2\theta$ from the
two and one generation cases, Eqs.~(\ref{onegen}) and (\ref{twogen}).
Again, we have the freedom to rotate the three zero mass eigenstates into
each other so the above form for $O$ is not unique.
However, as before, this is the choice which reduces to the unit matrix in the limit
$m/M\rightarrow 0$ and facilitates the extraction of the suppression parameter.
Complete diagonalization requires an additional multiplication by a matrix
of the form
\begin{equation}
O' = \left[ \begin{array}{cc} I & 0 \\ 0 & \mathcal{A}
             \end{array}
      \right]\;,
\end{equation}
so that
\begin{equation}
{O'}^\mathrm{T}
\left[ \begin{array}{cc} 0 & 0 \\ 0 & \mathcal{M}
        \end{array}
\right] O' =
\left[ \begin{array}{cc} 0 & 0 \\ 0 &
           \mathcal{A}^\mathrm{T}\mathcal{M}\mathcal{A}
        \end{array}
\right] =
\left[ \begin{array}{cc} 0 & 0 \\ 0 & \mathcal{D}
        \end{array}
\right]\;,
\end{equation}
where
\begin{equation}
\mathcal{D} =
\left[ \begin{array}{ccc} M_1 & 0 & 0 \\
                           0 & M_2 & 0 \\
                           0 & 0 & M_3
        \end{array}
\right]\;.
\end{equation}
The exact expression for $\mathcal{A}$ and the
eigenvalues $M_i$ ($i=1,2,3$) will depend on the values of
$(\alpha, \beta, \gamma)$.  However, note that
\begin{equation}
\mathcal{V} \equiv
O\,O' =
\left[ \begin{array}{cc} \mathcal{O}_{11} & \mathcal{O}_{12} \\
                          \mathcal{O}_{21} & \mathcal{O}_{22}
        \end{array}
\right]
\left[ \begin{array}{cc} I & 0 \\ 0 & \mathcal{A}
        \end{array}
\right] =
\left[ \begin{array}{cc} \mathcal{O}_{11} & \mathcal{O}_{12}\mathcal{A} \\
                          \mathcal{O}_{21} & \mathcal{O}_{22}\mathcal{A}
        \end{array}
\right]\;,
\label{OOPrime}
\end{equation}
so we see that $\mathcal{O}_{11}$, the upper left $3\times 3$ block of $\mathcal{V}=O O'$,
is not affected by what the actual form of $\mathcal{A}$ is.
Therefore, we can read off the common suppression factor for the 
$Z\nu_1\nu_1$, $Z\nu_2\nu_2$, and $Z\nu_3\nu_3$ vertices from our expression 
for $O$ and we find
\begin{equation}
1-\varepsilon
= \left(1-\frac{2}{3}\sin^2\theta\right)^2
\approx 1 - \frac{4}{3}\theta^2
\approx 1 - \frac{3m^2}{M^2}\;.
\label{threegensup}
\end{equation}
Again, this can be adjusted independently of the heavy mass eigenvalues.
Masses for the light eigenstates can be generated via minute perturbations to
our mass matrix as in the two generation case, but we omit the details.
Mixing among the charged leptons will cause $\nu_1$, $\nu_2$, and $\nu_3$ to mix:
\begin{eqnarray}
\left[ \begin{array}{c} \nu_e \\ \nu_\mu \\ \nu_\tau 
       \end{array}
\right]
= \mathcal{U}
\left[ \begin{array}{c} \nu_1 \\ \nu_2 \\ \nu_3 
       \end{array}
\right]\;,
\label{leptonmix}
\end{eqnarray}
where $\mathcal{U}$ is the unitary mixing matrix.
This mixing breaks the flavor universality of the suppression parameters.
We find that the suppression parameter for flavor $f=e,\;\mu,\;\tau$ is
given by
\begin{eqnarray}
\varepsilon_f
& = & \varepsilon\left|\mathcal{U}_{f1}+\mathcal{U}_{f2}+\mathcal{U}_{f3}\right|^2 \cr
& = & \varepsilon
\left[ 1 + 2\Re\left( \mathcal{U}^*_{f1}\mathcal{U}_{f2}
                    + \mathcal{U}^*_{f2}\mathcal{U}_{f3}
                    + \mathcal{U}^*_{f3}\mathcal{U}_{f1}
               \right)
\right]\;.
\label{flavorsupp}
\end{eqnarray}
Note that as in the two generation case, the sum of the suppression
parameters is independent of the mixing,
\begin{equation}
\varepsilon_e + \varepsilon_\mu + \varepsilon_\tau 
= 3\varepsilon
= \frac{9m^2}{M^2} \;.
\end{equation}

The mixing of the heavy states with the light states
is given by $\mathcal{O}_{12}\mathcal{A}$, so we need the explicit form
of $\mathcal{A}$.   Let us work out a few examples:
\begin{enumerate}
\item $(\alpha, \beta, \gamma)=(0,1,-1)$.
\begin{equation}
\mathcal{M} =
\left[ \begin{array}{ccc}
        0 & mt & -mt \\
        \phantom{-}mt & (M+2mt) & 0 \\
        -mt & 0 & -(M+2mt)
        \end{array}
\right]\;,
\end{equation}
\begin{equation}
\mathcal{A} =
\left[ \begin{array}{ccc}
        c_\phi^2-s_\phi^2 & \sqrt{2}s_\phi c_\phi & i\sqrt{2}s_\phi c_\phi \\
        -\sqrt{2}s_\phi c_\phi &  c_\phi^2 & -i s_\phi^2 \\
        -\sqrt{2}s_\phi c_\phi & -s_\phi^2 &  i c_\phi^2
        \end{array}
\right]\;,
\end{equation}
where
\begin{equation}
c_\phi = \cos\phi\;,\qquad
s_\phi = \sin\phi\;,\qquad
\tan2\phi \equiv \frac{ 2\sqrt{2}\,t^2 }{ 3 + t^2 }\;,
\end{equation}
and
\begin{equation}
\mathcal{A}^\mathrm{T}
\mathcal{M}
\mathcal{A} = \sqrt{M^2+6m^2}
\left[ \begin{array}{ccc}
        0 & 0 & 0 \\ 0 & 1 & 0 \\ 0 & 0 & 1
        \end{array}
\right]\;,
\end{equation}
\begin{eqnarray}
\mathcal{O}_{12}\mathcal{A}
& = & \frac{2}{3}sc
\left[ \begin{array}{ccc}
        (\cos 2\phi - \sqrt{2}\sin 2\phi) &
        (\cos 2\phi + \frac{1}{\sqrt{2}}\sin 2\phi) &
        i(\cos 2\phi + \frac{1}{\sqrt{2}}\sin 2\phi) \\
        (\cos 2\phi - \sqrt{2}\sin 2\phi) &
        (\cos 2\phi + \frac{1}{\sqrt{2}}\sin 2\phi) &
        i(\cos 2\phi + \frac{1}{\sqrt{2}}\sin 2\phi) \\
        (\cos 2\phi - \sqrt{2}\sin 2\phi) &
        (\cos 2\phi + \frac{1}{\sqrt{2}}\sin 2\phi) &
        i(\cos 2\phi + \frac{1}{\sqrt{2}}\sin 2\phi)
        \end{array}
\right]\;\cr
& = & \frac{m}{\sqrt{M^2+6m^2}}
\left[ \begin{array}{ccc}
        \cos 2\theta & 1 & i \\
        \cos 2\theta & 1 & i \\
        \cos 2\theta & 1 & i
        \end{array}
\right]\;.
\end{eqnarray}
This model includes a massless sterile neutrino while 
the massive states are degenerate.

\item $(\alpha,\beta,\gamma)=(1,1,-2)$.
\begin{equation}
\mathcal{M} =
\left[ \begin{array}{ccc}
        (M+2mt) & 2mt     & -mt \\
        2mt     & (M+2mt) & -mt \\
        -mt     & -mt     & -2(M+2mt)
        \end{array}
\right]\;,
\end{equation}
\begin{equation}
\mathcal{A} =
\left[ \begin{array}{ccc}
                   \frac{1}{\sqrt{2}} &
                   \frac{1}{\sqrt{2}} c_\varphi &
                   \frac{i}{\sqrt{2}} s_\varphi \\
                 - \frac{1}{\sqrt{2}} &
                   \frac{1}{\sqrt{2}} c_\varphi &
                   \frac{i}{\sqrt{2}} s_\varphi \\
                   0 & -s_\varphi & i c_\varphi
        \end{array}
\right]\;,
\end{equation}
\begin{equation}
\mathcal{A}^\mathrm{T}
\mathcal{M}
\mathcal{A} = 
\left[
\begin{array}{ccc}
M & 0 & 0 \\
0 & M_-  & 0 \\
0 & 0 & M_+
\end{array}
\right]\;,
\end{equation}
where
\begin{equation}
c_\varphi = \cos\varphi\;,\quad
s_\varphi = \sin\varphi\;,\quad
\tan 2\varphi \equiv \frac{ 4\sqrt{2}\,t^2 }{ 9 + 7\,t^2 }\;,\quad
M_\pm \equiv \frac{ 3\sqrt{M^2 + 8 m^2} \pm M }{ 2 } \;,
\end{equation}
and
\begin{eqnarray}
\mathcal{O}_{12}\mathcal{A}
& = & \frac{2}{3}sc
\left[ \begin{array}{ccc}
        0  &
         (\sqrt{2} c_\varphi - s_\varphi) &
        i(\sqrt{2} s_\varphi + c_\varphi) \\
        0  &
         (\sqrt{2} c_\varphi - s_\varphi) &
        i(\sqrt{2} s_\varphi + c_\varphi) \\
        0  &
         (\sqrt{2} c_\varphi - s_\varphi) &
        i(\sqrt{2} s_\varphi + c_\varphi)
        \end{array}
\right] \cr
& = & \frac{ \sqrt{6} m }{ \sqrt{ M_+^2 - M_-^2 } }
\left[ \begin{array}{ccc}
       0 & \sqrt{M/M_-} & i\sqrt{M/M_+} \cr
       0 & \sqrt{M/M_-} & i\sqrt{M/M_+} \cr
       0 & \sqrt{M/M_-} & i\sqrt{M/M_+}
      \end{array}
\right] \;.
\end{eqnarray}
Note that the three mass eigenvalues are different in this case and
that the heavy state with mass $M$ decouples from the light states.

\item $(\alpha,\beta,\gamma)=(1,\omega,\omega^2)$, $\omega^3=1$.
\begin{eqnarray}
\mathcal{M} & = &
\left[ \begin{array}{ccc}
        (M+2mt)    & -\omega^2 mt   & -\omega mt \\
      -\omega^2 mt & \omega (M+2mt) & -mt \\
      -\omega   mt & -mt            & \omega^2(M+2mt)
        \end{array}
\right]   \cr
& = &
\left[ \begin{array}{ccc}
        1 & 0 & 0 \\ 0 & \omega^2 & 0 \\ 0 & 0 & \omega
        \end{array}
\right]
\left[ \begin{array}{ccc}
        (M+2mt) & -mt & -mt \\
        -mt & (M+2mt) & -mt \\
        -mt & -mt & (M+2mt)
        \end{array}
\right]
\left[ \begin{array}{ccc}
        1 & 0 & 0 \\ 0 & \omega^2 & 0 \\ 0 & 0 & \omega
        \end{array}
\right]\;,
\end{eqnarray}
\begin{equation}
\mathcal{A} =
\left[ \begin{array}{ccc}
        1 & 0 & 0 \\ 0 & \omega & 0 \\ 0 & 0 & \omega^2
        \end{array}
\right]
\left[ \begin{array}{ccc}
        \frac{1}{\sqrt{3}} &  \frac{1}{\sqrt{6}} &  \frac{1}{\sqrt{2}} \\
        \frac{1}{\sqrt{3}} &  \frac{1}{\sqrt{6}} & -\frac{1}{\sqrt{2}} \\
        \frac{1}{\sqrt{3}} & -\frac{2}{\sqrt{6}} &  0
        \end{array}
\right]\;,
\end{equation}
\begin{equation}
\mathcal{A}^\mathrm{T}
\mathcal{M}
\mathcal{A} =
\left[ \begin{array}{ccc}
        M & 0 & 0 \\
        0 & \sqrt{M^2 + 9 m^2} & 0 \\
        0 & 0 & \sqrt{M^2 + 9 m^2}
        \end{array}
\right]\;,
\end{equation}
\begin{equation}
\mathcal{O}_{12}\mathcal{A}
= \sqrt{\frac{3}{2}}
\frac{ m\omega^2 }{ \sqrt{ M^2 + 9 m^2 } }
\left[ \begin{array}{ccc}
        0  &  -1 & i \\
        0  &  -1 & i \\
        0  &  -1 & i
        \end{array}
\right]\;.
\end{equation}
In this case, the heavy state of mass $M$ decouples from the
light states while the other two states have degenerate mass.

\end{enumerate}
As these examples illustrate, there is a wide range of possibilities
for the masses and mixings of the heavy states.

\section{The Dirac Masses and Quark--Lepton Unification}

We next consider the consequences of setting the Majorana masses
to zero in our model.   Our mass matrix will be
\begin{equation}
\left[ \begin{array}{cccccc}
        0 & 0 & 0 & \alpha m & \beta m & \gamma m \\
        0 & 0 & 0 & \alpha m & \beta m & \gamma m \\
        0 & 0 & 0 & \alpha m & \beta m & \gamma m \\
        \alpha m & \alpha m & \alpha m & 0 & 0 & 0 \\
        \beta  m & \beta  m & \beta  m & 0 & 0 & 0 \\
        \gamma m & \gamma m & \gamma m & 0 & 0 & 0
        \end{array}
\right]\;
\label{unificationtexture}
\end{equation}
which is manifestly rank 2. The non--zero eigenvalues are
\begin{equation}
\pm\;m\sqrt{|\alpha|^2+|\beta|^2+|\gamma|^2}\;.
\end{equation}
Therefore, this mass matrix leads to one massive, and two massless
Dirac fermions.

If we assume that the up-type quarks have the same Dirac mass
texture as the neutrinos, which would be the case in the
Pati-Salam \cite{Pati:uk} model, we obtain one massive
quark which can be identified with the $t$,
and two massless quarks which can be identified with the $u$
and the $c$.
Recall that a similar mechanism is at work in the
`democratic' \cite{democratic,democratic2} quark mass texture,
\begin{equation}
\left[ \; \bar{q}_1 \; \bar{q}_2 \; \bar{q}_3 \;\right]
\left[ \begin{array}{ccc} m & m & m \\
                          m & m & m \\
                          m & m & m \\
       \end{array}
\right]
\left[ \begin{array}{c} q_1 \\ q_2 \\ q_3
        \end{array}
\right]\;,
\end{equation}
which is manifestly rank 1, and has eigenvalues $0$, $0$, and $3m$,
which are identified as the $u$, $c$ and $t$ quark masses.
Thus, our model provides an alternative to the `democratic' model
to explain why the $t$ quark mass is so large compared to the other quarks.

For this idea to work, it must be the case that
\begin{equation}
m\sim 100\;\mathrm{GeV}\;,
\end{equation}
to predict the correct $t$ mass.
Since $\varepsilon\sim 0.003$ \cite{LOTW1} requires
\begin{equation}
\frac{m}{M} \sim 0.03\;,
\end{equation}
we can predict
\begin{equation}
M \sim 3\,\mathrm{TeV}\;.
\end{equation}
This places the heavy states within range of observation at the LHC.

\section{Constraint from $\mathbf{\mu\rightarrow e\gamma}$}

\begin{figure}
\begin{center}
\scalebox{0.75}{\includegraphics{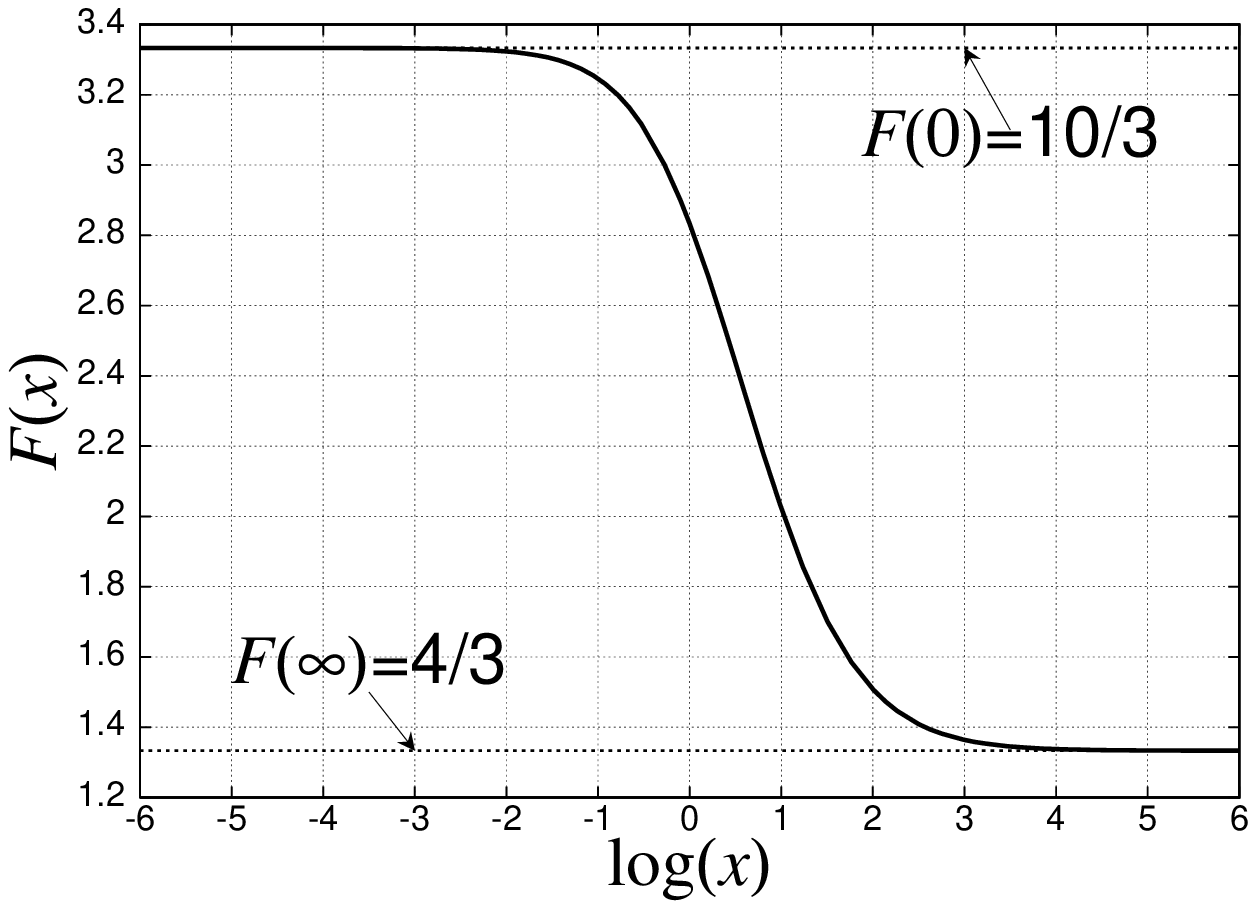}}
\caption{The behavior of the function $F(x)$.}
\label{Flogx}
\end{center}
\end{figure}

The flavor changing process $\mu\rightarrow e\gamma$ places
a strong constraint on our model.  The current limit on the
decay branching fraction is from the MEGA collaboration \cite{mega}
\begin{equation}
B(\mu\rightarrow e \gamma) 
= \frac{\Gamma(\mu\rightarrow e\gamma)}
       {\Gamma(\mu\rightarrow e\nu\bar{\nu})}
< 1.2 \times 10^{-11}\;,\qquad \mbox{(90\% confidence level)}\;.
\label{megalimit}
\end{equation}
The theoretical value of this branching fraction is given by \cite{mutoegamma}
\begin{eqnarray}
B(\mu\rightarrow e \gamma)
& = & \frac{3\alpha}{32\pi}
  \left| \sum_{i,j=1}^3 \mathcal{U}_{e i}^*\mathcal{U}_{\mu j} T_{ij} 
  \right|^2 \;, \cr
T_{ij} 
& = & \sum_{k=1}^6 \mathcal{V}_{ik}^* \mathcal{V}_{jk} F(m_k^2/m_W^2)\;,
\end{eqnarray}
where $\mathcal{U}$ is the mass diagonalization matrix for the charged leptons, 
Eq.~(\ref{leptonmix}), and $\mathcal{V}=OO'$ is the corresponding 
matrix for the neutrinos, 
Eq.~(\ref{OOPrime}).
$m_k$ is the mass of the $k$--th neutrino mass eigenstate,
and the function $F(x)$ is given by
\begin{equation}
F(x) = 2(x+2)I^{(3)}(x) - 2(2x-1)I^{(2)}(x) + 2xI^{(1)}(x)+1\;,
\end{equation}
in which
\begin{equation}
I^{(n)}(x) = \int_0^1 dz \frac{z^n}{z + (1-z)x}\;.
\end{equation}
$F(x)$ is plotted in Figure~\ref{Flogx}.
Using the unitarity of $\mathcal{V}$
we can express $T_{ij}$ as a sum over the heavy states only,
\begin{equation}
T_{ij} =
\sum_{k=\mathrm{heavy}}
\mathcal{V}_{ik}^* \mathcal{V}_{jk} \left[ F(m_k^2/m_W^2) - F(0) \right] \;,
\end{equation}
where the masses of the light states are all set to zero.
If we assume $m_\mathrm{heavy} \sim O(\mbox{TeV})$, it is evident 
from Figure~\ref{Flogx} that
\begin{equation}
F(m_\mathrm{heavy}^2/m_W^2) \approx F(\infty) = \frac{4}{3}\;.
\end{equation}
Then,
\begin{equation}
T_{ij} \approx 
-2\sum_{k=\mathrm{heavy}} \mathcal{V}_{ik}^*\mathcal{V}_{jk}\;.
\end{equation}

\begin{figure}
\begin{center}
\scalebox{0.75}{\includegraphics{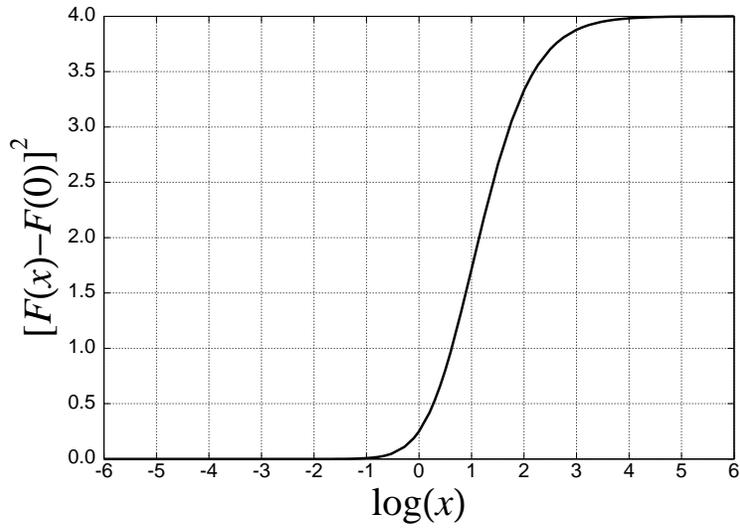}}
\caption{The $x$-dependence of $[F(x)-F(0)]^2$.}
\label{Flogx2}
\end{center}
\end{figure}

For the sake of simplicity, let us first consider the two generation case
we considered in section~III.
From Eq.~(\ref{twogenO}), we find
\begin{equation}
T_{ij} 
= -4s^2c^2
\left[ \begin{array}{cc} 1 & 1 \\ 1 & 1 \end{array} \right]
\approx -2\varepsilon
\left[ \begin{array}{cc} 1 & 1 \\ 1 & 1 \end{array} \right]\;.
\end{equation}
Then, taking $\mathcal{U}$ to be the mixing matrix in Eq.~(\ref{twogenmix}),
we obtain
\begin{equation}
B(\mu\rightarrow e\gamma)
= \frac{3\alpha}{32\pi}
  \left| 2\varepsilon \cos 2\phi \right|^2 
= \frac{3\alpha}{8\pi} \varepsilon^2 \cos^2 2\phi
= \frac{3\alpha}{8\pi} \varepsilon_e \varepsilon_\mu \;.
\end{equation}
Assuming $\varepsilon = 0.003$, we find
\begin{equation}
\frac{3\alpha}{8\pi}\varepsilon^2 
\approx 2\times 10^{-8}\;,
\label{PREFACTOR}
\end{equation}
which is more than three orders of magnitude above the MEGA limit,
Eq.~(\ref{megalimit}).  Note that had we assumed the heavy states
to be as light as the $Z$, then this prefactor would have been
suppressed by an order of magnitude as can be gleaned from
Figure~\ref{Flogx2}.  But even then, it would have been two orders
of magnitude above the experimental limit.
Thus, in order to suppress this process we need
\begin{equation}
\cos 2\phi \approx 0 \qquad\rightarrow\qquad
\phi \approx 45^\circ\;.
\end{equation}
This mixing breaks the flavor universality of the suppression
parameters maximally.

For the three generation model, the overall normalization of $T_{ij}$
depends on the number of heavy states that contribute to the sum.
For the three cases considered earlier, we find
\begin{equation}
T_{ij} = 
-2\eta\varepsilon
\left[ \begin{array}{ccc} 1 & 1 & 1 \\ 1 & 1 & 1 \\ 1 & 1 & 1 
       \end{array}
\right] \;,
\end{equation}
with
\begin{equation}
\eta =
\left\{ \begin{array}{ll}
        2/3 \;,\qquad & (\alpha,\beta,\gamma)=(0,1,-1) \;, \\
        1   \;,\qquad & (\alpha,\beta,\gamma)=(1,1,-2),(1,\omega,\omega^2) \;.
        \end{array}
\right.
\end{equation}
Note that in the $(\alpha,\beta,\gamma)=(0,1,-1)$ case, there are only
two massive states that contribute to $\eta$.
Thus,
\begin{equation}
\sum_{i,j=1}^3
\mathcal{U}_{ei}^*\mathcal{U}_{\mu j} T_{ij}
= -2\eta\varepsilon
\left[\,\mathcal{U}_{e1}
       +\mathcal{U}_{e2}
       +\mathcal{U}_{e3}
\right]^*
\left[\,\mathcal{U}_{\mu 1}
       +\mathcal{U}_{\mu 2}
       +\mathcal{U}_{\mu 3}
\right]\;,
\end{equation}
and
\begin{eqnarray}
B(\mu\rightarrow e\gamma)
= \frac{3\alpha}{8\pi}\eta^2\varepsilon^2
\left| \mathcal{U}_{e1}
     + \mathcal{U}_{e2}
     + \mathcal{U}_{e3}
\right|^2
\left| \mathcal{U}_{\mu 1}
     + \mathcal{U}_{\mu 2}
     + \mathcal{U}_{\mu 3}
\right|^2
= \frac{3\alpha}{8\pi}\eta^2
  \varepsilon_e\varepsilon_\mu\;.
\end{eqnarray}
The only difference from the two generation case is the factor $\eta^2$.
From this expression it is evident that to
suppress $B(\mu\rightarrow e\gamma)$ the mixing matrix must
be chosen to suppress either $\varepsilon_e$ or
$\varepsilon_\mu$.  
That is, we must identify one of the flavor eigenstates, 
$\nu_e$ or $\nu_\mu$, as a linear combination of
$\nu_1$, $\nu_2$, and $\nu_3$ which almost
completely decouples from the heavy states.

This result implies that in order to satisfy the experimental 
limit on $\mu\rightarrow e\gamma$,
the mixing angles in $\mathcal{U}$ must be generically large.
These mixing angles should not be confused with the mixing
angles in the MNS matrix \cite{Maki:mu} which are the angles
measured in neutrino oscillation experiments \cite{SK,SNO,K2K}.
The MNS matrix in our notation is given by
\begin{equation}
V_\mathrm{MNS} = \mathcal{U}\mathcal{V}_\mathrm{upper-left}\;,
\end{equation}
where $\mathcal{V}_\mathrm{upper-left}$ is the upper left
$3\times 3$ block of $\mathcal{V}$.  (Note that $V_\mathrm{MNS}$
is not unitary since $\mathcal{V}_\mathrm{upper-left}$ is but a
portion of an unitary matrix $\mathcal{V}$.)
Therefore, in order to make a connection to the experimental data
one must take into account the mixings among the light neutrinos 
in $\mathcal{V}$ also.
In the current model, as long as the light neutrinos are
exactly massless and degenerate, the MNS matrix is ill
defined.  The perturbations which break the degeneracy
and allow non-zero masses must be chosen judiciously to
obtain the correct $\mathcal{V}$ which would lead to
the correct MNS matrix.

\section{Conclusions and Discussion}

In this paper, we have shown explicit examples which demonstrate how 
large mixings between light active neutrinos and heavy sterile states 
can be realized within the seesaw framework,
\textit{i.e.} with the same number of active and sterile neutrinos,
by a judicious choice of the mass texture.
This required the presence of more than one generation so that there 
exist sufficient degrees of freedom in the mass texture to permit independent 
tuning of the mixing angles and mass eigenvalues \cite{Chang:1994hz}.
Indeed, even though the mass textures we considered, 
Eqs.~(\ref{twogentex}) and (\ref{threegentex}), were constrained 
to a generation universal form, there were still enough degrees of 
freedom left to accomplish this objective.

Due to the independence of the mass eigenvalues and the mixing angles
in these models, the only experimental constraint on the masses of the 
heavy states is that they must be heavier than the $Z$ \cite{LOTW1}.
However, if we invoke quark--lepton universality and require the 
up--type quarks to share the same Dirac mass texture as the neutrinos, 
then we find that:
\begin{enumerate} 
\item it provides a non--democratic explanation as to why the $u$ and $c$ quarks are
almost massless compared to the $t$ quark,
\item the Dirac mass scale must be about 100 GeV to produce the correct $t$ mass, and
\item the heavy sterile states must have mass of a few TeV to produce the correct
suppression parameter $\varepsilon$.
\end{enumerate}
This last observation stands in stark contrast to 
Grand Unified Theory (GUT) scenarios in which the masses of the 
sterile states are at the GUT scale, far exceeding experimental reach.   
If the heavy states are indeed in the few TeV range, they could be observed at the
LHC. Of course, this estimation is from one particular model, and we cannot
discount the possibility that the mass is actually much lower and could be 
observed at the Tevatron Run II at Fermilab.

Enforcing generation universality among the light neutrinos necessarily 
reduces the rank of our mass matrix leading to zero mass eigenvalues for 
the light states. To obtain non-zero masses, one must raise 
the rank by breaking generation universality with small perturbations.
The same is true of the quark sector also since neither the $u$ nor the 
$c$ is massless.  It is interesting to ask whether the same 
universality breaking perturbation could generate the correct mass ratios 
and mixings for both the neutrino and up--quark sectors.

Our model comfortably accomodates the introduction
of down quark-charged lepton mass unification in addition to the
up quark-neutrino unification utilized above.  The constraints from
$\mu\rightarrow e\gamma$ mandate large mixing angles among the charged
leptons in our model.  In models with down quark-charged lepton 
unification ({\it e.g.}, the Pati-Salam model) this implies large
mixing angles in the down quark mass matrix.  For the CKM matrix to 
have the small mixings that are observed, the large down quark mixings
must be cancelled by large up quark mixings.  
The presence of large mixings in the neutrino
Dirac mass matrix then follows from up quark-neutrino unification and 
is consistent with the texture of Eq.~(\ref{unificationtexture}).
Models of quark-lepton unification which generate the
large mixing angles of the MNS matrix using large angles in the 
charged lepton mass matrix have been explored extensively \cite{largeangle}.
Such models generically relate the MNS and CKM matrices.  In the present
context, however, explicit prediction of the CKM and MNS matrices would 
require a detailed delineation of the small perturbations of the 
mass matrices, which is beyond the scope of this paper.

Due to the typically large mixings between the light and heavy states
in our model, the process $\mu\rightarrow e\gamma$ provides a strong
constraint.  
We have found that the product $\varepsilon_e\varepsilon_\mu$ must
be strongly suppressed.  
Since either $\varepsilon_e$ or $\varepsilon_\mu$ must remain sizable
to account for the discrepancy between the $Z$-pole and NuTeV data,
lepton flavor universality among the 
suppression parameters $\varepsilon_\ell$, ($\ell = e,\mu,\tau$),
must be violated.  Lepton universality is 
constrained experimentally at the 
$1\%$ level by $W$ decays \cite{pich},
\begin{eqnarray*}
\left|\frac{g_\mu}{g_e}\right| &=& 1.000 \pm 0.011 \\
\left|\frac{g_\tau}{g_\mu}\right| &=& 1.026 \pm 0.014 \\
\left|\frac{g_\tau}{g_e}\right| &=& 1.026 \pm 0.014, 
\end{eqnarray*}
and at the $0.2\%$ level by $\tau$ decays \cite{pich},
\begin{eqnarray*}
\left|\frac{g_\mu}{g_e}\right| &=& 1.0006 \pm 0.0021 \\
\left|\frac{g_\tau}{g_\mu}\right| &=& 0.9995 \pm 0.0023 \\
\left|\frac{g_\tau}{g_e}\right| &=& 1.0001 \pm 0.0023. 
\end{eqnarray*}
Whether lepton universality and 
$\mu \rightarrow e \gamma$ constraints can be satisfied
simultaneously within our model while maintaining the fit to 
the $Z$-pole and NuTeV data will be addressed in a subsequent paper
\cite{nextone}.

The model analyzed in this paper should be contrasted with the recent
analysis of neutrino masses and mixings by Glashow \cite{Glashow:2003wt}.
In his work, he assumes, as we do, that neutrino masses arise from a
seesaw mass texture involving three active and three sterile neutrino states.
He postulates a particular neutrino mass texture with the property that
the mixings between light and heavy states are non-universal, leading
to non-universal suppressions. In particular, the suppression of the
neutrino weak gauge boson couplings occurs in a single light--mass eigenstate.
The large mixing angles in the MNS matrix \cite{Maki:mu,SK,SNO,K2K}
spread the suppression among the three flavors and enhance the 
$\mu\rightarrow e\gamma$ rate, providing a limit on the size of the
suppression parameter $\varepsilon$.
In fact, Glashow's model is similar to ours, although he uses
a different basis for the mass matrix.  His analysis starts from a 
basis where the inter-generational mixing is minimal.
In our analysis, we have taken the opposite approach.
We start from a basis in which the inter-generational mixing is maximal
and the suppression of light neutrino couplings to the weak gauge bosons 
are universal. 
To suppress the $\mu\rightarrow e\gamma$ rate, we must require large
mixings among the charged lepton flavors leading to the violation of
flavor universality of the suppression parameters.
In both approaches, the violation of flavor universality
is essential for obtaining agreement with current experimental data.
It is amusing to note that lepton flavor universality violation is 
\textit{required} for lepton flavor conservation.

Given the strong constraint on our model from the flavor changing
process $\mu\rightarrow e\gamma$, one is lead to question what constraints
will be provided by other flavor changing processes such as 
$\mu$--$e$ conversion in nuclei.  This is a particularly interesting process
since the MECO (Muon to Electron COnversion) experiment at Brookhaven \cite{meco}
proposes to improve upon the current limits on $\mu$--$e$ conversion \cite{PDB:2002} 
by more than three orders of magnitude.  The MEG (Mu-E-Gamma) experiment at
PSI \cite{muegamma} also proposes to improve upon the MEGA limit on 
$\mu\rightarrow e\gamma$ by about two orders of magnitude.
We will investigate the implications of the potential MECO and MEG 
limits in a future publication.

Finally, we propose that if heavy neutral particles 
such as those discussed in this paper are found to exist they be 
called ``neutrissimos'' instead of ``heavy sterile neutrinos'' or
``heavy neutral leptons'', the terms commonly used in the literature.  
Those names are balky, and the first one is misleading, 
since the particles are neither completely ``sterile'' nor are 
they ``-ino'' compared to
the neutrons.

\section*{Acknowledgments}

We would like to thank Sandor~Benczik, Lay~Nam~Chang, Naoyuki~Haba,
Yasuhiro~Okada, and Alexey~Pronin for helpful discussions.
We would also like to thank Massimiliano~Di~Ventra for coining the term
`neutrissimo'.
This research was supported in part by the U.S. Department of Energy,
grants DE--FG05--92ER40709, Task A (T.T. and N.O.),
and DE-FG02-84ER40153 (L.C.R.W.).


\newpage

\end{document}